\documentclass[12pt]{iopart}
\usepackage{amssymb}
\usepackage{graphicx}
\begin{document}
\title{\bf Efficient Degradation and Expression Prioritization 
with Small RNAs}
\author{Namiko Mitarai$^{1,2}$,
Anna M. C. Andersson$^1$, \\
Sandeep Krishna$^1$, 
Szabolcs Semsey$^{1,3}$, 
and Kim Sneppen$^1$ 
}
\address{
$^1$ Niels Bohr Institute,
Blegdamsvej 17, DK-2100, Copenhagen, Denmark.}
\address{
$^2$Department of Physics, Kyushu University 33,
Fukuoka 812-8581, Japan.}
\address{
$^3$Department of Genetics,
Eotvos Lorand University, Budapest H-1117, Hungary.}
\ead{namiko@stat.phys.kyushu-u.ac.jp}

\begin{abstract}
We build a simple model for feedback systems involving
small RNA (sRNA) molecules based on the iron metabolism system
in the bacterium {\sl E. coli}, and compare it with the corresponding system
in {\sl H. pylori} which uses purely transcriptional regulation.
This reveals several unique features of sRNA based regulation
that could be exploited by cells.
Firstly, we show that sRNA regulation can maintain a smaller turnover
of target mRNAs than transcriptional regulation, without
sacrificing the speed of response to
external shocks.
Secondly, we propose that a single sRNA can
prioritize the usage of different target mRNAs.
This suggests that sRNA regulation would be
more common in more complex systems which need to
co-regulate many mRNAs efficiently.
\end{abstract}
\submitto{\PB}
\section{Introduction}
Small noncoding RNAs (sRNAs) have
recently been discovered
as key components of genetic regulation
in systems ranging from bacteria to 
mammals~\cite{G04,FGJLJLBB05,JBB06},
and this has spurred much
activity in understanding their functional advantages.
For example, degradation of sRNA with their target mRNAs has been
proposed as a mechanism to obtain
ultrasensitivity \cite{LMLKWB04}.

One particular system which has
been extensively studied is the Fe-Fur system in the bacterium
{\sl Escherichia coli} which contains the regulatory
sRNA RyhB~\cite{ARR03,MA05}.
This system is responsible for maintaining homeostasis of
Fe$^{++}$ ions, which are essential for cell functioning
but also poisonous at high concentrations.
During aerobic exponential growth, iron-using enzymes in {\sl E. coli}
utilize around $10^6$ Fe atoms
per cell generation, but more than
$10^4$ Fe$^{++}$ ions in free or loosely-bound form
is poisonous \cite{NOBOMKY99,KI96}.
Thus, the cell faces the problem of maintaining a huge flux
of Fe through a small reservoir, and at the same time
channeling this flux into its most essential functions
when the cell is starved of iron:
It needs to prioritize and sort the usage of a limited resource.

\begin{figure}[t]
\begin{center}
\includegraphics[width=8cm]{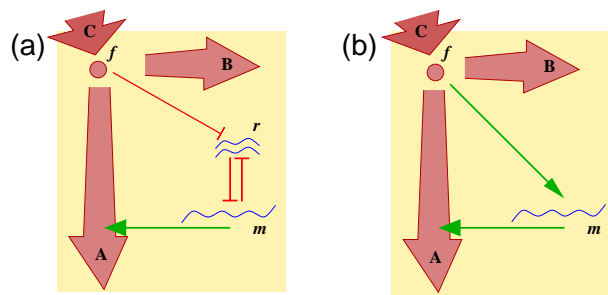}
\end{center}
\caption{(a) The sRNA motif
found in {\sl E. coli}, consists of
three variables, $f$, the
Fe-Fur complex which depends on the amount of loosely bound iron,
$r$, the sRNA ryhB, and $m$,
the mRNA of iron-using proteins.
The red barred arrows between $r$ and $m$ represent
the formation of the $r-m$ complex and its subsequent
degradation. The other red barred arrow indicates
transcriptional repression of $r$ by $f$.
The influx of $f$, denoted by $C$, is divided into
the channel $A$, regulated by $m$ (green arrow),
and the non-regulated channel $B$.
(b) The transcription motif found in {\sl H. pylori}.
Here, $m$ is transcriptionally activated by $f$.
\label{motifs}} 
\end{figure}

Reference \cite{SAKJMS06} describes a detailed model of the Fe
regulation in {\sl E. coli},
which incorporates several feedback mechanisms that
together secure the system against both
up and down shifts of the iron level.
In this paper,  we focus on the
prioritization of the usage of iron
by the sRNA regulation and its role in sudden
iron depletion. 
The model in \cite{SAKJMS06} is simplified
into a core ``motif" (Figure ~\ref{motifs}a)
describing the negative feedback used in iron homeostasis.
The motif consists of three variables,
$f$ the iron-activated Fur (Ferric uptake
regulator) protein complex
that senses the Fe$^{++}$ level, $r$, the sRNA RyhB, and $m$,
the mRNA of iron-using proteins.
The sRNA works by binding strongly to the mRNA, after which this
entire complex is rapidly degraded \cite{MG02,MEG03,MVG05}.
The sRNA is in turn transcriptionally repressed by $f$.
Thus, $f$ effectively activates $m$, through a double negative link
via the sRNA.

Interestingly, the regulation of iron homeostasis in the bacterium
{\sl Helicobacter pylori} differs from that of {\sl E. coli} in one
important respect: the regulation via the sRNA RyhB is replaced
by a direct transcriptional activation of the mRNA $m$ by $f$
(see Figure~\ref{motifs}b) \cite{DSRS01,DRDCRS06}.
These two bacteria motivate us to compare
the ``sRNA motif" with the corresponding ``transcription motif".

By studying these motifs, we demonstrate 
the following interesting aspects of sRNA regulation:
(i) both motifs can be adjusted to have similar response times,
but the metabolic cost is different for the two motifs.
(ii) a single type of sRNA can
efficiently prioritize expression level of various 
downstream target mRNAs, 
thus prioritizing the usage of a limiting resource.
We also discuss possible experiments to test these results.

\section{The Two Motifs}
\subsection{sRNA motif}

In the motif of Figure~\ref{motifs}a we focus
on the case where sRNA 
binds to mRNA and both RNAs in the complex are degraded.
In terms of an effective rate constant for the 
overall degradation, $\delta$, the dynamics of the concentrations of 
the sRNA ($r$) and its target mRNA ($m$) 
can be described by
\begin{eqnarray}
\frac{dr}{dt} & = & \alpha_r - \frac{r}{\tau_r} - \delta \cdot r \cdot m\\
\frac{dm}{dt} & = & \alpha_m - \frac{m}{\tau_{mrna}} - \delta \cdot r
\cdot m
\end{eqnarray}
where $\alpha_r$ and $\alpha_m$ set the respective production rates,
and $\tau_r$ and $\tau_m$ define the background degradation 
times.
We next rescale the parameters 
by measuring concentrations in units
of $\alpha_m \tau_r$ and measuring time in units of $\tau_r$
\footnote{The equation are formally rescaled by replacing
$m\rightarrow m/(\alpha_m \tau_r )$,
$r\rightarrow r/(\alpha_m \tau_r )$
and $t\rightarrow t/\tau_r$. Thereby, a unit of time corresponds to the
degradation time of the sRNA, and the unit of production rates
corresponds to the production rate of the target mRNA.}.

To these rescaled equations we also add
an equation for $f$, a small-molecule-activated
transcription factor for the sRNA (as shown in 
Figure~\ref{motifs}a).
We simplify this two step reaction from $f$ to regulation 
by assuming that all bindings are first order, and by rescaling
binding constants such that $f=1$ results in half-repression
of the promoter of the sRNA gene.
Including the import and consumption of $f$
in analogy to the Fe-Fur system\cite{SAKJMS06}, 
the full dynamics of the motif becomes:
\begin{eqnarray}
\frac{df}{dt} &=&
\left\{
\begin{array}{ll}
C- A \cdot m\;
- B \cdot  f  \;\;\; & \mbox{when}\;\; f>0,
\label{fs}
\\
C \;\;\; & \mbox{when} \;\; f=0,
\end{array}\right. \\
\frac{dr}{dt} &=& \frac{\alpha}{1+f}- r
- \gamma  r\cdot m \label{rs},
\\
\frac{dm}{dt} &=&1\; -\; \frac{m}{\tau_m}
- \gamma  r\cdot  m.\label{ms}
\end{eqnarray}
Here the dimensionless mRNA degradation time 
$\tau_m=\tau_{mrna}/\tau_r$, and the  
dimensionless degradation rate of the RNAs is related
to the dimensionfull parameters as follows:
\begin{equation}
\gamma = \delta \alpha_m \tau_r^2.
\end{equation}

In the absence of $m$, $r$ is
degraded relatively slowly because unbound sRNA are quite 
stable {\sl in vivo} \cite{MEG03}.
This means that $\tau_r$, which is 
our rescaled time unit in eqs. (\ref{fs})-(\ref{ms}), is 
around $(25/\ln 2)$ min.
In the absence of $r$, $m$ is degraded with $\tau_m=0.2$, 
the lifetime of RyhB target
mRNA in {\sl E. coli} from ref. \cite{MEG03}.

\begin{figure}[t]
\begin{center}
\includegraphics[angle=-90,width=8cm]{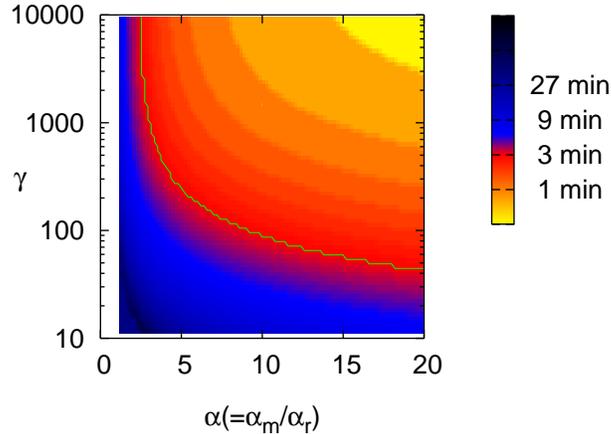}
\end{center}
\caption{
Limits on rescaled sRNA-target mRNA coupled degradation rate
$\gamma=\delta \alpha_m \tau_r^2$. 
The colour shading indicate 
that the model takes longer (shorter) time 
in a darker (yellow) region to be depleted 5-fold
after the small RNA is activated by setting $f$ from 
40 to zero at $t=0$.
The green solid line shows the contour line of 3 min. 
When production of mRNA is much faster than
production of sRNA the 5-fold reduction can be obtained by a small $\gamma$,
whereas a relatively small $\alpha=\alpha_m/\alpha_r$ makes production
of sRNA so slow that the mutual degradation must be very fast.
The response time is calculated based on the 
degradation time of sRNA $\tau_r=(25/\ln 2)$min \cite{SAKJMS06}. 
\label{parameters}} 
\end{figure}

The most important parameters that 
determine the dynamics of sRNA regulation are $\alpha$ and $\gamma$
in eqs.(\ref{rs}) and (\ref{ms}).
We can determine a reasonable range of values for
these two parameters using experimental data.
First, for the Fe-Fur system $\alpha$ and $\gamma$
are mutually constrained by the observation that the target mRNA, sodB,
is depleted around 5-fold within 3 minutes after 
full induction of the RyhB promoter\cite{MEG03}. These 3 minutes
include the time required for RyhB production, set by $\alpha$, plus 
the time for the produced RyhB to 
bind to sodB message and to degrade the complex, set by $\gamma$.
For $\alpha<1$ there will never be enough RyhB to deplete
the message completely, whereas for $\alpha$ slightly higher than 1
an extremely high $\gamma$ is needed for efficient depletion.

Fig. \ref{parameters} illustrates this.
We simulate eqs. (\ref{rs}) and (\ref{ms}) 
to measure how long it takes for $m$ to be depleted 5-fold 
after fully-activating sRNA at $t=0$
by changing $f$ manually from $40$ to zero.
The solid line in Fig. \ref{parameters} 
shows values of $\gamma$ and $\alpha$ that 
give a 3 minutes depletion-time for sodB mRNA.
With $\alpha$ values in the range from 3 to 10, we see that 
physiological $\gamma$ values would need to be between 100 and 10000.
In addition, the reduction of iron consumption occurs
as soon as the sRNA-mRNA complex is formed, even before
the mRNA is degraded, but
the measurement of \cite{MEG03} does not 
distinguish RNA species in complexes from the free form.
Thus, the effective $\gamma$ could be larger than the above estimate.
The value of $\alpha$ can be estimated from other
data in ref. \cite{MEG03} of 
RyhB and sodB time series after induction and 
repression of RyhB.
From these data, we estimate that $\alpha$
is between 2 and 5.
In the rest of the paper, we
explore the small RNA motif 
for a range of $\alpha$ and $\gamma$ values,
around the estimates made above.

The three other parameters, $A$, $B$, and $C$,
in the iron flux equation (\ref{fs}), we set
using experimental data \cite{SAKJMS06},
as described in the appendix.

\subsection{Transcriptional motif}

To model the transcription motif (Figure~\ref{motifs}b) we
replace (\ref{rs}) and (\ref{ms}) by the single equation:
\begin{equation}
\frac{dm}{dt} 
= D \; \cdot \frac{f^h}{f^h+K_t^h} \; -\; \frac{m}{\tau_m}.
\label{mt}
\end{equation}
The first term models the direct transcriptional
activation of $m$
production by $f$, where $D$ sets the
maximum production rate of $m$ and $K_t$ sets
the binding constant between $f$ and the DNA.
The ``Hill coefficient" $h$ sets
the steepness of the response, and is related to the
cooperativity in binding.
We use the same values of the influx $C$ and
two constants $A$ and $B$ in (\ref{fs})
as those used in the sRNA motif.
We choose $D$ and $K_t$ in the transcription
motif to have the identical steady state values of the $f$ and $m$
to the corresponding sRNA motif (parameterized by $\alpha$ and $\gamma$)
for high iron ($f=40$) and low iron ($f=5$).
It is not obvious that this is
possible for all values of $\alpha$ and $\gamma$.
In fact, we found that it is not possible
for $h\le 2$, but
when $h=3$, we can set $D$ and $K_t$
such that the above conditions are fulfilled for $\alpha\in[0,20]$ and
$\gamma\in [0,2000]$. Therefore, henceforth we keep $h=3$.
\begin{figure}
\begin{center}
\includegraphics[angle=-90,width=13cm]{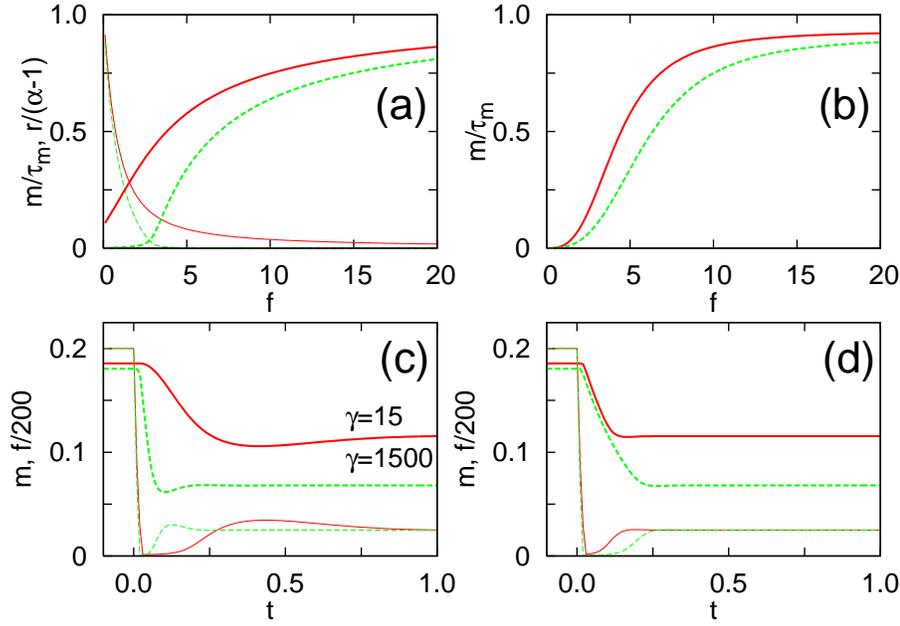}
\end{center}
\caption{(a) The steady state level of 
$r/(\alpha-1)$ 
(thin lines) and $m/\tau_m$ (thick solid lines) vs. $f$ for $\alpha=4$ with
$\gamma=15$ (red solid lines) and $\gamma=1500$ (green dashed lines)
in the sRNA motif.
(b) The steady state level
of $m/\tau_m$ (thick lines) vs. $f$ for the
corresponding transcription motif parameters (see text).
The time evolution of $f$ (divided by 200,
thin lines) and $m$ (thick lines)
for the systems in (a) and (b)
are shown in (c)
and (d), respectively, after
the sudden drop of $C$ at $t=0$.
}
\label{ss_and_dyn}
\end{figure}

\section{Results}
\subsection{Degradation times}

Figures~\ref{ss_and_dyn}a 
and b show how the steady state values
of $r$ and $m$ depend on the steady state $f$ level,
in the sRNA motif for two different $\gamma$ values, 
and, respectively, in the corresponding
transcription motif systems.
For the sRNA motif, a larger $\gamma$ results
in a much larger ratio between
maximum and minimum $m$ values, and a steeper drop between 
them (Figures~\ref{ss_and_dyn}a).
This is expected because in the $\gamma\rightarrow\infty$
limit, $m = \max(O(1/\gamma),\tau_m[1-\alpha/(1+f)])$,
and this steepness in the sRNA regulation
is referred to as ``ultrasensitivity'' in \cite{LMLKWB04}.
For the transcription motif, on the other hand,
the dependence of $m$ on $f$ is weaker
and nearly unaffected by the corresponding change
in $D$ and $K_t$ (Figure~\ref{ss_and_dyn}b).
This is because the steepness of the curve in the transcription 
motif is determined by the Hill coefficient $h$,
which is set to be relatively high value 3 but 
still not enough to give as sharp slope as the 
sRNA motif with $\gamma=1500$.

The importance of $\gamma$ is also evident
in the dynamical response of the motifs
to a sudden depletion in the external
iron source $C$ (Figures~\ref{ss_and_dyn}c and d).
For the sRNA motif, Figure~\ref{ss_and_dyn}c shows that a larger $\gamma$
results in a faster drop in $m$ level, and a quicker approach to the
new steady state level. The same is true for the $f$ level also.
That is, a large $\gamma$ naturally ensures a faster removal of all excess $m$,
while also allowing $f$ and
$m$ to climb back to non-zero steady levels even after $f$ drops almost to zero
during the initial shock.
In contrast, the transcription motif displays approximately
the same timescale of mRNA drop for the two corresponding cases
because a drop in $m$ takes
a time proportional to $\tau_m$, independent of other parameters.
\begin{figure}
\begin{center}
\includegraphics[width=8cm]{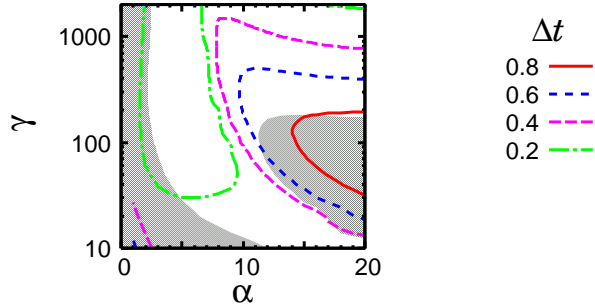}
\end{center}
\caption{The contour plot of sRNA motif response time, $\Delta t$,
at which $f$ recovers to and subsequently remains within
95\% of the final steady state value.
The corresponding transcription motif systems show a
faster response in the shaded region.
\label{contour}
 }
\end{figure}

We investigate this further by quantifying the response time.
Looking at Figure~\ref{ss_and_dyn}c, we see that the $f$ level drops
very sharply and then rises towards its final steady state value.
During this rise, at some time, $\Delta t$, $f$ reaches within
95\% of the final steady state value. We use $\Delta t$ as a measure of the response speed
of the system. In Figure~\ref{contour} we show a contour plot
of $\Delta t$ for a range of values of $\alpha$ and $\gamma$.
The corresponding transcription motif systems show a faster
response than the sRNA motif in the shaded region.
The comparison indicates that sRNA based translational regulation
produces a faster response than transcriptional regulation
when $\alpha\gtrsim 2$ (and $\alpha$ is not too large
compared to the investigated range of $f$)
and $\gamma$ is larger than a critical level, around 150
(the unshaded region in Figure~\ref{contour}).

\subsection{Metabolic cost}

The relatively slow response of the transcription motif is, of
course, because its response timescale is set by $\tau_m$, which we
keep constant. We emphasize that it is possible
to achieve a fast response in the transcription motif also by
decreasing $\tau_m$. There are some costs associated with this
alternative strategy though. Faster mRNA turnover due to lower
$\tau_m$ requires a higher production to maintain the same
homeostatic levels of $f$. At high $f$ levels, where mRNA is high,
the sRNA motif secures a low degradation rate of mRNA, whereas the
transcription motif produces the mRNA at its maximum rate. On the
other hand, at low $f$ the sRNA motif maintains a high rate of mRNA
degradation, while the transcription motif saves resources by
reducing mRNA production. Therefore, for a given response speed to
sudden iron starvation, the sRNA motif is less costly if the
bacterium usually lives in iron-rich  conditions, whereas the
transcription motif is preferable if the organism mostly lives in
iron-poor conditions.

\subsection{sRNA prioritize downstream protein levels}
\subsubsection{Prioritization in the iron feedback motif}
\begin{figure}
\begin{center}
\includegraphics[angle=-90,width=13cm]{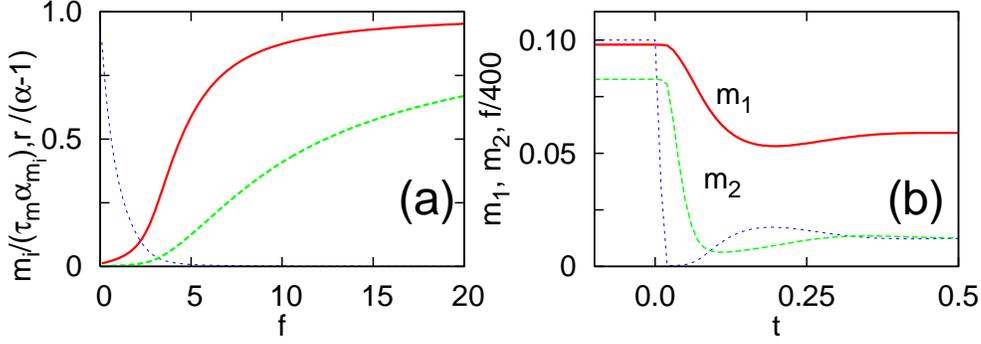}
\end{center}
\caption{
Investigation of mRNA prioritization by the sRNA motif.
(a) The steady state level of 
$r/(\alpha-1)$
(blue dotted line), $m_1/(\tau_m\alpha_{m_1})$
(red solid line), and $m_2/(\tau_m\alpha_{m_2})$ (green dashed line)
vs. $f$ for $\alpha=4$, $\gamma_1=150$,
and $\gamma_2=1500$.
(b) The time evolution of $f$ (divided by 400, blue dotted line),
$m_1$ (red solid line), and $m_2$ (green dashed line)
after the sudden drop of $C$ at $t=0$.
\label{sorting}
}
\end{figure}

The properties of sRNA regulation can be
used in an interesting way when more than one kind
of mRNA is under regulation \cite{MVG05}.
Different mRNA can have
hugely different binding strengths to the regulating sRNA, and thereby
very different effective degradation timescales. 
This is because the
binding strength of the $r$-$m$ complex is, to a first
approximation, an exponential
function of the number of matching base-pairs, which can vary by
an order of magnitude across different mRNA:
The free-energy gain per matching 
base-pair is around 1 to 2 kcal/mol,
which can give the difference in 
statistical weight $\exp(\Delta G/k_B T)\approx 5$ to 30. 
Because of this property,
sRNA regulation could be used to prioritize the degradation
of different mRNA.
We illustrate this by adding a second mRNA to the sRNA motif:
\begin{eqnarray}
\frac{dr}{dt} &=&
\frac{\alpha}{1+f}- r
- r  \cdot \left(\gamma_1 m_1+\gamma_2 m_2\right)
\label{2rs}
\\
\frac{dm_1}{dt}&=&\alpha_{m_1}-\frac{m_1}{\tau_m}
- \gamma_1  r  m_1,\label{2m1s}\\
\frac{dm_2}{dt}&=&\alpha_{m_2}- \frac{m_2}{\tau_m}
- \gamma_2  r  m_2. \label{2m2s}
\end{eqnarray}
For $f$, we use (\ref{fs}),
replacing $m$ by $(m_1+m_2)$.
The two mRNAs, $m_1$ and $m_2$, have different effective
degradation rates, $\gamma_1$ and $\gamma_2$, resulting in
different steady state levels for a given $C$ value. Other parameters
are set as in the case with a single mRNA.
Figure~\ref{sorting} shows the
behavior of this system. 
The steady state level in Figure~\ref{sorting}a
shows that $m_2$, the mRNA with larger $\gamma$,
is suppressed more than $m_1$ on depletion of $f$.

The prioritization behavior is easily understood by considering the extreme
case where both $\gamma$s are very large but $\gamma_2\gg\gamma_1$.
Then, depending on the level of $f$, the steady state is such that
either (i) both mRNAs are near-zero, (ii) only $m_1$ is non-zero,
(iii) both $m_1$ and $m_2$ are non-zero.

Taking into account finite $\gamma$ values, the prioritization 
efficiency becomes, respectively,
\begin{itemize}
\item[(i)]
$m_1\approx O(1/\gamma_1)$ and $m_2\approx O(1/\gamma_2)$ for small
$f$, i.e. $f\lesssim \frac{\alpha}{\alpha_{m_1}+\alpha_{m_2}}-1$,
\item[(ii)]
$m_1\approx \tau_m(\alpha_{m_1}+
\alpha_{m_2}-\frac{\alpha}{1+f})$ and $m_2\approx
O(\gamma_1/\gamma_2)$ for intermediate $f$, i.e.
$\frac{\alpha}{\alpha_{m_1}+\alpha_{m_2}}-1\lesssim f\lesssim
\frac{\alpha}{\alpha_{m_2}}-1 $,
\item[(iii)]
$m_1\approx \tau_m \alpha_{m_1} $ and $m_2\approx
\tau_m(\alpha_{m_2}-\frac{\alpha}{1+f})$ for large $f$, i.e.
$\frac{\alpha}{\alpha_{m_2}}-1 \lesssim f$.
\end{itemize}
where the $O(x)$ are some functions that
are small and proportional to $x$ for small $x$.

Case (ii) is what we refer to as the ``prioritized state", where only one of
the mRNA (the one with smaller $\gamma$) is present and the other's
level drops to near-zero. Clearly, the larger the difference between
the $\gamma$ values of the mRNAs, the better the prioritization; 
the order of magnitude difference in $\gamma$ is important 
to have the clear prioritization.
In addition, $\alpha$, the sRNA production rate, determines the range
of $f$ for which the sRNA is affected, and a larger $\alpha$ results
in prioritization being effective for a wider range of $f$. 

In addition, the dynamics in Figure~\ref{sorting}b shows that,
when $C$ is suddenly dropped,
the mRNA with a larger $\gamma$ value is rapidly depleted while the
other mRNA stays at a higher level.
That is, the sRNA is not only able to 
prioritize the mRNA steady
state levels, but is also able to remove the ``unwanted'' mRNA $m_2$
much quicker than the ``wanted'' mRNA $m_1$.

\subsubsection{Prioritization of multiple mRNAs}
\begin{figure}
\begin{center}
\includegraphics[angle=-90,width=7cm]{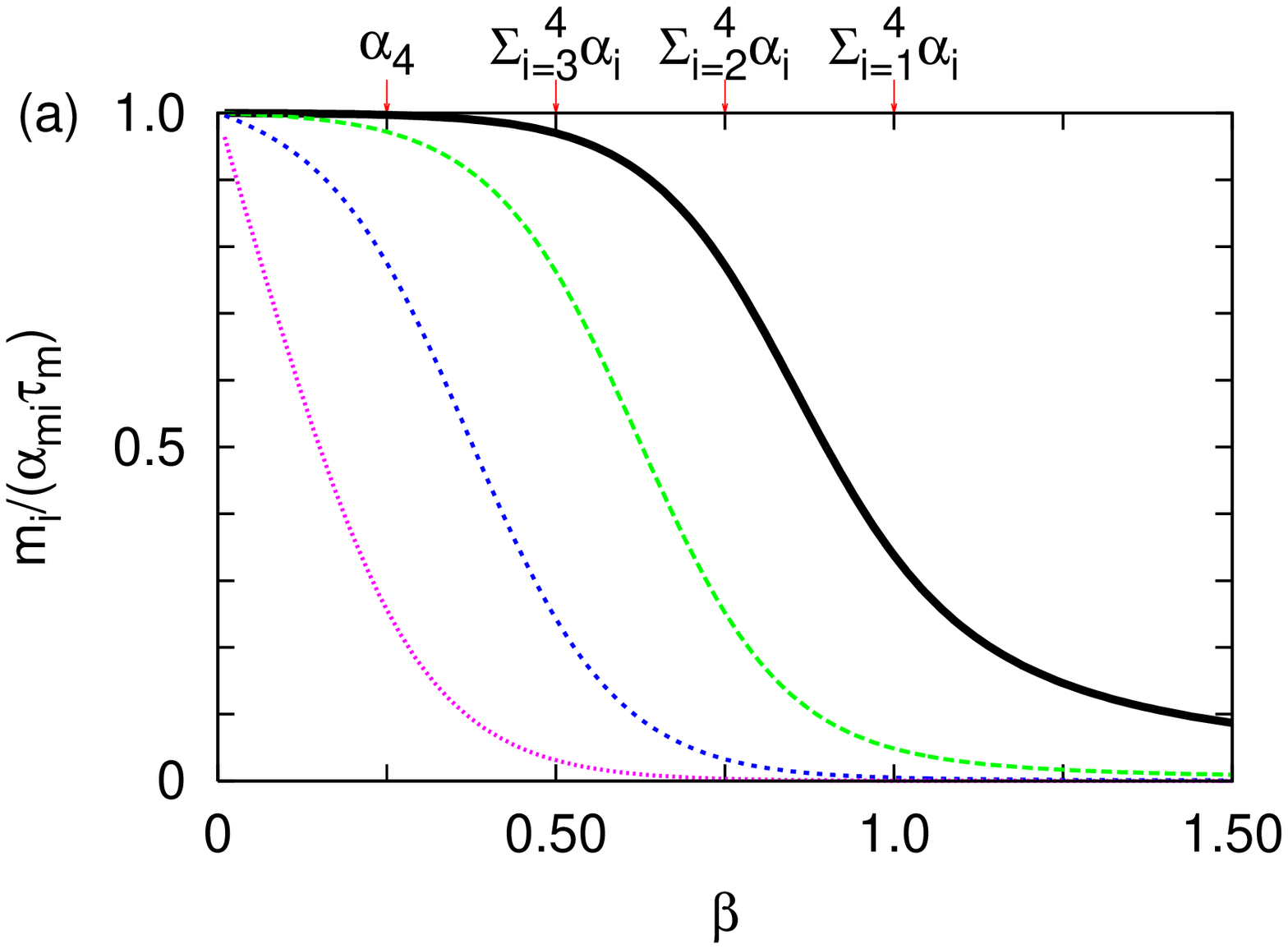}
\includegraphics[angle=-90,width=7cm]{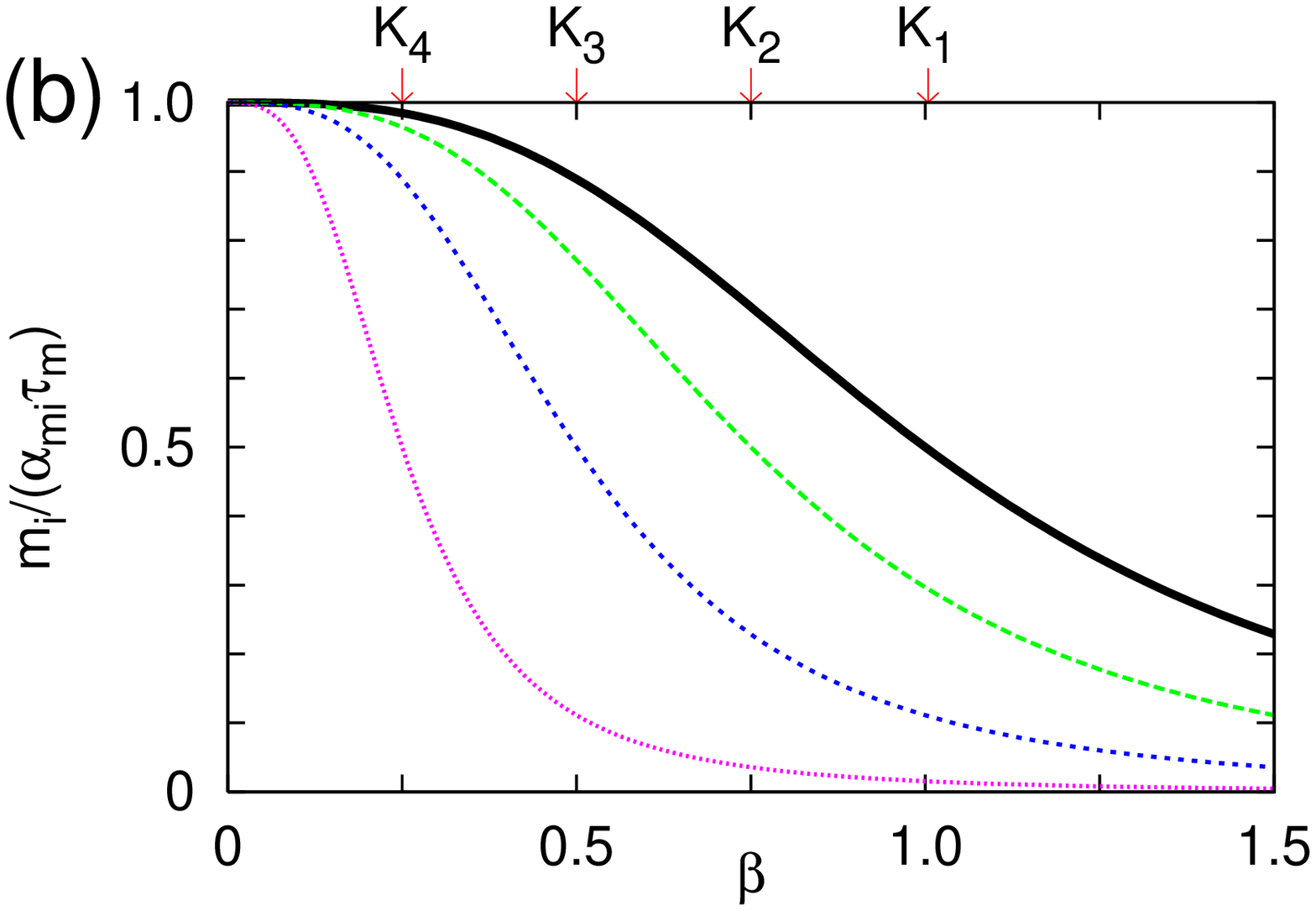}
\end{center}
\caption{
Prioritization of mRNAs by (a) sRNA regulation 
and (b) transcriptional regulation.
The steady state level of mRNAs normalized by their maximum value 
is plotted against $\beta$. 
(a) The prioritization by sRNA regulation, for
$\alpha_{m_1}=\alpha_{m_2}=\alpha_{m_3}=\alpha_{m_4}=0.25$,
and $\gamma_1=10^2, \gamma_2=10^3,\gamma_3=10^4,\gamma_4=10^5$.
The value of $\gamma_i$ are chosen to be large and 
have 10-fold difference between different mRNAs, so that 
the separation becomes clear. 
(b) Transcriptional regulation, where the transcription factor 
has concentration $\beta$ and acts as a repressor.
The Hill coefficient $h$ is 3.
The binding constants are 
$K_1=1$, $K_2=0.75$, $K_3=0.5$, and $K_4=0.25$. 
The values of $K_i$ are chosen to be
$K_i=\sum_{j=i}^4 \alpha_i$, so that 
the value of $\beta$ at which the mRNA starts to be significantly degraded 
is the same for the two motifs.
}
\label{SortingSimple}
\end{figure}
It is clear that the prioritization can be generalized to the case of 
more than two kinds of mRNAs regulated by a single type of sRNA. 
To illustrate this we ignore feedback through $f$ and consider the 
following general system with $n$ different types of mRNAs:
\begin{eqnarray}
\frac{dr}{dt}&=&\beta-r-\sum_{i=1}^n \gamma_i r m_i,
\label{mrss}\\
\frac{dm_i}{dt}&=&\alpha_{mi}-\frac{m_i}{\tau_m}-
\gamma_i  r m_i 
\quad \mbox{for}\quad i=1,n.
\end{eqnarray}
Here, to focus on the  prioritization, the production term of 
sRNA in (\ref{2rs}) which  contains
feedback from iron concentration is replaced by a constant $\beta$.
We assume
$\gamma_1<\gamma_2<\cdots<\gamma_i<\gamma_{i+1}<\cdots<\gamma_n$
without loss of generality,
and rescale all variables to be dimensionless 
such that $\sum_{i=1}^n\alpha_{m_i}$ and
the degradation time of sRNA are unity.

\begin{figure}
\begin{center}
\includegraphics[width=8cm]{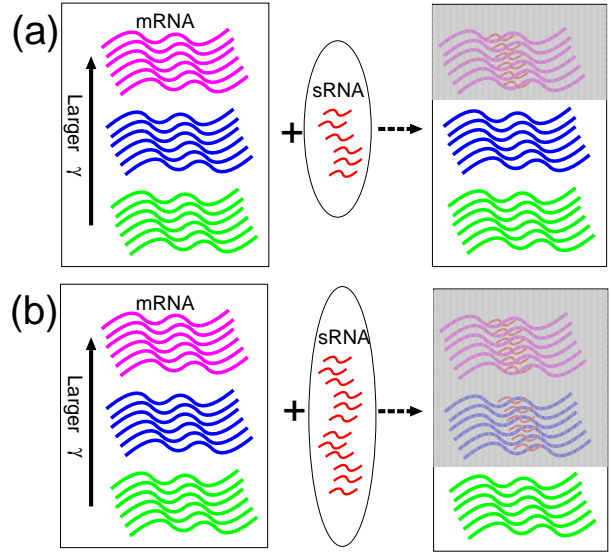}
\end{center}
\caption{Schematic picture of the prioritization of 
various kinds of mRNAs by a single type of sRNA. 
Different kinds of mRNAs are represented by different
colors, and have different value of $\gamma$ (left boxes). 
(a) When a small amount of sRNAs is produced, the mRNAs with larger 
$\gamma$ are degraded (shaded region in the right box), 
while the levels of mRNAs with small $\gamma$
are scarcely affected.
(b) As more sRNA is produced, 
the mRNAs with smaller $\gamma$ are also degraded.
}\label{sortfig}
\end{figure}

Figure \ref{SortingSimple}a shows the normalized 
steady state level of mRNAs, $m_i /(\alpha_{m_i}\tau_m)$,
versus the production rate of the sRNA, $\beta$.
As $\beta$ becomes larger, sRNA increases and more
mRNAs are degraded.  As shown in Figure \ref{SortingSimple}a,
the $n$-th mRNA with the largest $\gamma$ is degraded
first. This also ``protects'' other mRNAs from degradation because 
the sRNAs are also degraded together with
the $n$-th mRNAs. The ($n-1$)-th mRNA starts to be 
degraded when the level of $n$-th mRNA becomes low enough,
which occurs roughly at $\beta\approx \alpha_n$.
The ($n-2$)-th mRNA starts to be degraded when $\beta\approx
\alpha_{n}+\alpha_{n-1}$, and so on, and 
finally all the mRNAs are almost completely 
degraded when $\beta\approx \sum_{i=1}^n\alpha_{i}=1$.
The separation of the level between the ($k+1$)-th mRNA 
and the $k$-th mRNA for 
$\sum_{i=k+1}^{n}\alpha_{mi}<\beta<\sum_{i=k}^{n}\alpha_{mi}$
becomes clearer for larger 
difference between $\gamma_{i+1}$ and $\gamma_i$%
\footnote{
In the case $\gamma_{i+1}/\gamma_i\ll 1$ for any $i$,
the steady state level of mRNA for 
$\sum_{i=k+1}^n\alpha_{m i}<\beta<\sum_{i=k}^n\alpha_{m i}$
is estimated as follows:
$m_{i}/(\tau_m \alpha_{mi})\approx 1$ for $i<k$,
$m_{k}/(\tau_m \alpha_{mk})
\approx (\sum_{i=k}^n\alpha_{m i}-\beta)/\alpha_{mk}$,
and $m_i/(\tau_m\alpha_i)\ll 1$ for $i\ge k+1$.
}.
This multistep-switch-like degradation 
upon changing the value of $\beta$ 
is the characteristic feature of the sRNA regulation. 
This prioritization mechanism is schematically 
presented in Figure \ref{sortfig}, where the mRNAs are 
degraded in descending order of the value of $\gamma$. 

We note that it is possible to 
have different steady state of mRNAs in the transcriptional 
motif also.
Suppose $r$ is the concentration of the 
repressor with a production rate $\beta$. 
Then the simplest model for the mRNA level is
\begin{eqnarray}
\frac{dr}{dt}&=&\beta-r,\\
\frac{dm_i}{dt}&=&\frac{\alpha_{m_i}}{1+(r/K_i)^h}-\frac{m_i}{\tau_m}
\quad \mbox{for}\quad i=1,n.
\end{eqnarray}
with a Hill coefficient $h$ and binding constants $K_i$;
we assume $K_1>K_2>\cdots >K_n$ without loss of generality.
Here, the variables are rescaled to be dimensionless such that 
$K_1$ and the degradation time of $r$ are unity.
The normalized steady state level is given by
$m_i/(\alpha_{m_i}\tau_m)=\frac{1}{1+(\beta/K_i)^h}$:
The levels decrease as $\beta$ increases with the slope 
determined by the Hill coefficient $h$, and the characteristic 
value of $\beta$ where the $m_i$ level becomes half of its 
maximum is given by $K_i$.
Figure~\ref{SortingSimple}b shows the
separation of the steady state by transcription regulation 
with a relatively high value of the Hill coefficient $h=3$ 
and various values of $K_i$.
We see that the separation of the $m_i$ level 
is not as sharp as with sRNA regulation,
and $m_i$ does not change much upon changing $\beta$
especially for large $K_i$.
The sRNA regulation is more effective 
in the sense that the prioritization of the mRNAs 
is sensitive to small changes in $\beta$.
\section{Discussion}
Our analysis pinpoints three features that are particular to sRNA
regulation in a feedback system. First, as sRNAs act through
degradation, the regulation can, in principle, be very fast, 
generating a near instant response. Second, as the sRNA motif
uses a double negative link, instead of a direct 
activation regulation, it has a higher metabolic cost for conditions where
downstream targets are repressed. Third, and most interestingly, it
has the capability of efficiently prioritizing the usage of downstream target
genes.

It is essential for the prioritized 
expression of downstream targets that the sRNA as
well as the mRNAs are degraded together after they form a complex.
In the large $\gamma$ limit, the effective prioritization occurs because
the degradation of $m_2$ interferes with the degradation of $m_1$ by
sequestering $r$, leaving less unbound $r$ to bind $m_1$.
Indeed, if the sRNA simply catalyzed the degradation of the mRNAs without
itself being degraded (i.e., no $\gamma_i m_i$ terms in
(\ref{2rs}) or (\ref{mrss})), the ``protection" of $m_1$ is also lost. 
The degradation of different mRNAs doesn't interfere 
as in the previous case, and thus the prioritization 
is less effective.

The switch-like behavior due to the
``ultrasensitivity'' of the sRNA regulation \cite{LMLKWB04}
together with the prioritization suggested in this work opens
the possibility of more sophisticated regulation of gene 
expression. In particular, if each mRNA is targeted by 
several kinds of sRNAs, the combinatorics allows 
one to realize various logic gates.
For example, if different kinds of sRNAs 
can bind to the same part of the targeted mRNA
to trigger the regulation, it behaves as an ``OR'' gate.
Such an example is known in {\sl V. cholerae}, 
where four different sRNAs regulate HapR and 
any one of them is enough for regulation\cite{LMLKWB04}.
It is also, in principle, possible that one mRNA has multiple binding sites
for different sRNAs \footnote{
The authors do not know of an established example,
but there exists an mRNA that has multiple binding sites
for the same micro RNA in eukaryotic cells \cite{H04,CJFLH04}}. 
In this case, if binding of all the sRNAs is 
necessary to trigger the regulation, it realizes an ``AND'' gate.
The concentration dependent prioritization 
can add more 
complicated functions to the logic gates.
It is an interesting future issue to explore the possibility of 
combinatoric regulation by sRNAs.

We have tested numerically that the results of this paper do not
strongly depend on the specific form of (\ref{fs}) for $f$. As
long as in- and out-fluxes are large enough to allow a much faster
response of $f$ than of $m$, then $m$ is the rate-limiting factor.
Another thing to note is that we assumed that the Hill coefficient
for the repression of sRNA by $f$ is 1, though we introduced a Hill
coefficient $h=3$ in the transcription motif to achieve the sharp
contrast in $m$ at high and low $f$. If we put a Hill coefficient
$H>1$ in the sRNA motif by replacing the production term of $r$ with
$\alpha/(1+f^H)$, the $m$ vs. $f$ curve becomes even steeper, which
makes the response sharper.

\section{Experimental tests}

The suggested prioritization possibility for downstream mRNAs invites
experimental tests. One key quantity of interest is the degradation
parameter $\gamma$, which in fact sets the efficiency of the whole
sRNA regulatory system. For large $\gamma$, the sRNA
regulation works as a step function: when production of sRNA is
larger than production of downstream targets, these are instantly
removed. Therefore it is essential to measure degradation times of
downstream targets under various expression levels of the sRNA.
These degradation times are tightly coupled to the steady state
level of downstream mRNA, and could therefore be obtained from bulk
measurements, using for example microarrays for both RyhB and
downstream targets. For a given $r$ (=sRNA) in the steady state, the
downstream mRNA level is $m_i(r)=\alpha_{m_i}/(1/\tau_m+\gamma_i
r)$, and therefore the slope of $m_i(0)/m_i(r)-1$ versus $r$ gives
$\gamma_i \tau_m$.

An experimental test of the prioritization capability of the RyhB system is
to consider homeostasis, and compare wild type with any mutant where
the RyhB binding part of a downstream target gene is highly
over-expressed. The over-expressed genes should be constructed such
that they do not produce proteins that can bind Fe,
and thereby indirectly influence the free/loosely bound Fe pool. 
For a given low level of external iron, the prioritized usage implies 
that for mutants where the over-expressed gene 
has small $\gamma$ the expression levels of other genes
would be almost the same as in wild type.
For the remaining, there will be large changes associated to
RyhB being depleted by the over-expressed gene with large $\gamma$. 
If the difference in $\gamma$ between different target mRNA 
is of an order of magnitude, i.e., enough to have clear 
prioritization, the influence 
of the over-expressed gene should be quite sharp.
Further, our prioritization scheme implies that as the external Fe 
is depleted further, the number of downstream genes 
which influence homeostasis should diminish monotonically.
\section{Conclusion}

Using the well characterized homeostatic response system for iron in
bacteria, we analyzed the pros and cons of sRNA versus transcription
regulation. The investigated negative feedback motif of
Fe $\rightarrow$ FeFur $\rightarrow$ proteins $\rightarrow$  Fe brings out
the functional similarities and differences between the two
alternative strategies of regulating downstream targets. For
sufficiently high Hill coefficients, transcriptional regulation can
reproduce the same steady state behavior as the sRNA regulation.
Further, both regulations can in principle provide a fast response
to sudden decreases in externally available iron. However, their
functional capabilities differ in two important aspects.

\begin{itemize}
\item
First, adjusting parameters to obtain similar response times, the
sRNA motif results in more turnover of target mRNAs in iron-poor
conditions, whereas the transcription motif results in more
turnover in iron-rich conditions.

\item
Second, the sRNA allows a prioritization of expression level of
downstream targets, thus efficiently regulating the usage of a limiting
iron resource. At the same time, unwanted mRNA is degraded more rapidly.
This observation fits with the fact that the
transcription motif is found in {\sl H. pylori}, a bacterium with a
small genome and limited capacity for genome regulation, while {\sl
E. coli}, which has a larger genome and can benefit from fine tuning
of mRNA levels, has an sRNA motif.
\end{itemize}

Our analysis suggests new ways to analyze other systems where
multiple sRNAs regulate a more complicated response, including for
example quorum sensing~\cite{LMLKWB04}. There, mutual binding inhibition
between the sRNAs and a central regulator (LuxR mRNA in {\sl V.
harveyi} \cite{LMLKWB04}, HapR mRNA in {\sl V. cholerae} \cite{JH97},
and the translational regulator RsmA in Pseudomonas \cite{PWHHHCHW01}),
may allow signals from adhesion and host factors to differentiate
the sRNAs and thereby their downstream targets. Overall, our
analysis suggests that sRNAs or micro RNAs 
may allow a near-digital reorganization
of cellular composition, an observation which concurs with their
ubiquity in regulatory processes associated with development.\\

\ack
This work was funded by the Danish National
Research Foundation. NM thanks the Yamada Science Foundation for
supporting her stay at the NBI. SS is grateful for the Janos Bolyai
Research Fellowship of the Hungarian Academy of Sciences.

\section*{Appendix:
Parameters in the flux equation}
We set the parameters $A$, $B$ and $C$ in eq. (3)
using data on the uptake and usage of Fe in 
{\sl E. coli} and {\sl H. pylori}.
These systems are characterized by a huge
flux of free $Fe^{++}$, where the fluxes $C$, $A$ and $B$ are so large that
the pool is replenished about 100 times per cell generation.

In more detail, 
the constant incoming flux, $C$, is partitioned into two channels, A
and B, that are motivated by the two separate ways of using iron in
the Fe-Fur system. Flow through channel A is regulated, i.e., it can
be reduced during iron-starvation, and is proportional to the mRNA
level, but independent of $f$ 
when there is any substantial amount of $f$ in the system 
~\cite{SAKJMS06}%
\footnote{
When we simulate eq.(\ref{fs}),
we used the form $df/dt=C-A\cdot m\cdot 
[f/(f+K_{cut})] -B\cdot f$ 
with $K_{cut}=0.1$ to avoid the numerical difficulty 
due to the singularity at $f=0$. This expression 
agrees with eq.(\ref{fs}) in the limit of $K_{cut}\to 0$.
}. 
In contrast,
flow through channel B is proportional to $f$. In the case of the full
Fe-Fur system~\cite{SAKJMS06} the regulation of this flux by other
proteins becomes important when extracellular iron increases
suddenly, but here we focus on the regulation by sRNA and therefore
keep flow through channel B unregulated. That is, our motif is
designed to respond to depletion of $f$. Our ``minimal" motif cannot
do without this B channel because removing it results in robust
oscillations, for which there is no evidence in {\sl E. coli}
\cite{ARR03}.

The parameters in (\ref{fs}) are set 
as follows \cite{SAKJMS06}:
For conditions where extracellular iron is plentiful, we demand that
the internal Fe$^{++}$ level is such that $f=40$
\cite{SAKJMS06} in steady state,
while at the same time the net in- and out-flux
per cell generation is approximately 100 times larger,
representing the fast turnover
and high usage of Fe \cite{ARR03,SAKJMS06}.
In addition, we assume that the iron flux
is equally partitioned between
the A and B channels,
because we found it best fulfills the
homeostatic requirements
in our full model~\cite{SAKJMS06}.
These conditions set the value of two parameters
$C=4572$ (i.e. $C=114\times f$~\cite{SAKJMS06})
and $B=57$.
The value of $A$ is completely determined by
the steady state level of $m$, which depends on the parameters
in the regulation part of the motif 
(i.e., $\alpha$ and $\gamma$ in the sRNA motif,
or $D$ and $K_t$ in the transcription motif).
For iron-starvation we demand that $f=5$ \cite{SAKJMS06},
keeping the value of $B$ 
and $A$ constant.
This condition is achieved by reducing $C$,
reflecting the reduction in extracellular iron.
The $C$ change needed to get $f=5$ is dependent on
the values of the regulation part of the motif.
Note that, for a given regulatory motif, 
the steady state value of $f$ is 
a unique function of the influx $C$.


\end{document}